\documentclass[aps,pre,subeqns.floatfix,amsmath,twocolumn]{revtex4}
\usepackage[pdftex]{graphicx}
\usepackage{amsmath}
\usepackage{amsfonts}
\usepackage{graphicx}
\usepackage{hyperref}
\usepackage{mathtools}
\usepackage{verbatim}
\usepackage[outdir=./]{epstopdf}
\usepackage{subfigure}
\usepackage{hyperref}
\usepackage{latexsym}
\usepackage{dcolumn}
\usepackage{epsf}
\usepackage{float}

\usepackage{color}
\usepackage[utf8]{inputenc}

\begin{document}

\title{Perfect synchronization in complex networks with higher order interactions}
\author{Sangita Dutta$^1$}
\author{ Prosenjit Kundu$^2$} 
\email{jitprosen.math@gmail.com}
\author{Pitambar Khanra$^3$} 
\author{Chittaranjan Hens$^4$}
\author{Pinaki Pal$^1$}
\email{pinaki.pal@maths.nitdgp.ac.in}
\affiliation{$^1$Department of Mathematics, National Institute of Technology, Durgapur~713209, India}
\affiliation{$^2$Dhirubhai Ambani Institute of Information and Communication Technology, Gandhinagar, Gujarat, 382007, India}
\affiliation{$^3$Department of Mathematics, University at Buffalo, State University of New York, Buffalo, USA}
\affiliation{$^4$Center for Computational Natural Science and Bioinformatics, International Institute of Informational Technology, Gachibowli, Hyderabad 500032, India}
\begin{abstract}
We propose a framework for achieving perfect synchronization in complex networks of Sakaguchi-Kuramoto oscillators in presence of higher order interactions (simplicial complexes) at a targeted point in the parameter space. It is achieved by using an analytically derived frequency set from the governing equations. The frequency set not only provides stable perfect synchronization in the network at a desired point, but also proves to be very effective in achieving high level of synchronization around it compared to the choice of any other frequency sets (Uniform, Normal etc.). The proposed framework has been verified using scale-free, random and small world networks. In all the cases, stable perfect synchronization is achieved at a targeted point for wide ranges of the coupling parameters and phase-frustration. Both first and second order transitions to synchronizations are observed in the system depending on the type of the network and phase frustration. The stability of perfect synchronization state is checked using the low dimensional reduction approach.
The robustness of the perfect synchronization state obtained in the system using the derived frequency set is checked by introducing a Gaussian noise around it.  
\end{abstract}

\maketitle
\section{Introduction}
Higher order structures, such as three- and four-way interactions in addition to pairwise interactions, are widespread in neurological, biological, ecological, and sociological systems\cite{majhi2022dynamics, battiston2020networks, petri2021topological, alvarez2021evolutionary,iacopini2019simplicial,martignon1995detecting,yu2011higher,giusti2016two,reimann2017cliques,sizemore2018cliques}. In ecological systems such higher order interactions (HOI) where three or more species interact with each other can stabilize large ecological communities \cite{chatterjee2022controlling,grilli2017higher,bairey2016high}. On the other hand, recent studies suggest the mesoscopic organization of brain through higher order interactions allows efficient   information processing and offers useful guidelines for performing complex tasks \cite{petri2014homological,giusti2015clique,sizemore2018cliques,ganmor2011sparse}. Motivated by  these practical implications, the network science community is also concentrating  on understanding the various types of collective behavior ranging from synchronization to epidemic spreading that can go beyond conventional paired interactions \cite{battiston2020networks, petri2021topological, gambuzza2021stability, skardal2019abrupt}. 

\par
Synchronization observed in the flocking pattern of birds, or rhythmic flashing  of fireflies, can be modelled with the interacting nonlinear dynamical units \cite{Strogatz_synchronization_book,Pikovsky_synchronization_book,Kuramoto_chemical_book,Newman_network_book,Cohen_complex_book,barrat2008dynamical}. The classic Kuramoto dynamics, one of the celebrated dynamical model used for studying the synchronization \cite{kuramoto1984chemical, Acebron_RMP2005},  encodes the phase evolution of each node. In a complex network, nodes with non-identical natural frequencies are connected to one another via paired linkages. Depending on the coupling configuration or particular frequency design, such complicated pairwise connections could cause the entire system to reveal ``continuous", ``discontinuous", or ``optimal" synchronization \cite{kuramoto1984chemical, gomez2011explosive, leyva2012explosive, gomez2007paths, Zhang_PRL2015, coutinho2013kuramoto,ichinomiya2004frequency, skardal2019abrupt}. However, a slight phase-lag between the oscillators may destabilize or erode the stable synchronization states, and destroy the switch-like discontinuous synchronization transition for certain cases \cite{chamlagai2022grass,kundu2018perfect, kundu2017transition,kundu2020optimizing,kundu2019synchronization,sakaguchi1986soluble,skardal2015erosion}. 
Against this backdrop, we explore the impact of HOI in a network of coupled  phase-lag oscillators. Particularly, in presence of HOI, and fixed coupling strength, we seek a suitable frequency set which can lead the system to a {\it perfect}  \cite{kundu2018perfect} global synchronization state where all phases are in unison. 

The HOI can induce abrupt synchronization transitions between incoherent state and coherent state without any correlations between network structure and dynamical functions. These synchronized states are stable even in the presence of repulsive pairwise coupling due to such higher order interaction \cite{skardal2020higher}.  
 In a system, three-way interactions in addition to pairwise interactions can cause abrupt desynchronization transitions without any abrupt synchronisation transitions  and {extensive multistable partially synchronized states\cite{skardal2019abrupt}} may appear.

 Recent advances suggest that, higher order interaction may be vital and play an important role in general oscillator systems like brain dynamics~\cite{petri2014homological, giusti2016two, sizemore2018cliques}, collaboration networks~\cite{patania2017shape}, social contagion network~\cite{iacopini2019simplicial} etc. In particular, interaction in 2- simplexes are important to describe correlation in neuronal activity in brain~\cite{cui2017htm} providing a missing link between the structure and dynamics. In fact, considering higher order interaction is useful where different types of correlation exists between the nodes in a coupled oscillator systems.  Despite these findings, enhancement of synchronization in phase frustrated dynamics with higher order interaction has not been explored so far.

Recently, it has been demonstrated that a trade-off between pairwise and higher order interactions can result in a higher level of synchronization (optimal synchronization) in relatively weaker coupling strength \cite{skardal2021higher} by using an optimal frequency  set obtained from the dominant eigenvector of the composite Laplacian, which encodes both pairwise and triangular interactions in the absence of phase frustration. 
However, in spite of the reported nontrivial effects of phase frustration on the dynamics of pairwise networked system~\cite{sakaguchi1986soluble, omel2012nonuniversal, kundu2017transition, kundu2019synchronization, kundu2018perfect, brede2016frustration},  the role of it is least explored in networks in presence of higher order interactions.

Thus, in this paper, we focus on the topic of targeted
global synchronization in phase frustrated complex networks in presence of pairwise as well as higher order interactions and ask the question {\it `Can we determine a set of frequencies to achieve perfect synchronization at a targeted point of the parameter space in such networks?’} In the process, we employ the analytical approaches reported in~\cite{skardal2014optimal,kundu2018perfect} and develop a framework based on linear theory to achieve the goal of perfect synchronization at a targeted point. Analytical approach eventually leads to the determination of a set of frequencies involving the structural and dynamical properties of the network for the achievement of perfect synchronization. Detailed numerical demonstration with different networks following the analytic finally confirms the achievement of targeted perfect synchronization using the derived frequency.  

The organization of the paper is as follows. Sections~\ref{Model} and~\ref{Analytical} describe the model and the analytical frame work for deriving a frequency set for achievement of perfect synchronization at a targeted point. Subsequent section~\ref{Numerical} presents the detailed numerical results on perfect synchronization in different networks. Then the stability and robustness of the synchronization states are presented in the sections~\ref{lowd} and ~\ref{robustness}. A general discussion and conclusions are made in the section~\ref{discussion}. We proceed with the model description in the next section.

\section{Model Description}\label{Model}

We consider different networks of coupled Sakaguchi-Kuramoto~\cite{sakaguchi1986soluble,Kuramoto_chemical_book} phase oscillators of size $N$ with higher order interactions along with pairwise interactions~\cite{skardal2020higher,skardal2019abrupt,adhikari2022synchronization}. Dynamics of each oscillators in the network is governed by the equation
\begin{eqnarray}
\label{Kuramoto_model}
\dot{\theta_i}&=&\omega_i+ K_1 \sum_{j=1}^{N}A_{ij}\sin(\theta_j-\theta_i-\alpha)  \nonumber \\     
&&+\frac{K_2}{2} \sum_{j=1}^N \sum_{l=1}^N B_{ijl} \sin(2\theta_j-\theta_l-\theta_i-\beta), \\
&&\hspace*{4cm} i=1,2,\dots,N,   \nonumber
\end{eqnarray}
where $\theta_i$ is the phase and $\omega_i$ is the intrinsic natural frequency of the $i$-th oscillator. $\alpha$ and $\beta$ act as the frustration terms in the system corresponding to the pairwise and the triadic interactions respectively.  $K_1$ and $K_2$ are the coupling strengths associated with $1$-simplex (pairwise interaction) and $2$-simplex (triadic interaction) respectively. $A_{ij}$ is the $ij$-th element of the adjacency matrix $A=(A_{ij})_{N\times N}$ associated to the 1-simplex, where $A_{ij}= 1$, if $i$-th and $j$-th node are connected and $A_{ij}=0$, otherwise. Similarly, $B_{ijl}$ is the adjacency tensor associated to $2$-simplex where $B_{ijl}=1$ if there is a triadic connection between the $i$th, $j$th and $l$th nodes and $B_{ijl}=0$, otherwise. Note that the networks under consideration is undirected. As a result, we have $A_{ij}=A_{ji}$ and $B_{ijl}=B_{ilj}=B_{jil}=B_{lji}=B_{lij}=B_{jli}$ for all admissible $i$ and $j$. Here we choose $2$ simplicial complex, which leads to the consideration of the triadic and pairwise interactions only. By the definition of simplicial complex, every link in that triangle will be in the simplicial complex. So, the adjacency tensor can be written in the form $B_{ijl}=A_{ij}A_{jl}A_{li}$.  

To quantify the level of synchronization in the network we use the Kuramoto order parameter $r$ given by 
\begin{equation}
\label{order_parameter}
re^{i\psi(t)}=\frac{1}{N}\sum_{j=1}^{N}e^{i \theta_{j}},
\end{equation}
where $\psi(t)$ is the average phase of the oscillators at time $t$. The value of the order parameter lies between $0$ and $1$. $r=0$ indicates the system is in the incoherent state, while $r=1$ indicates the fully synchronized state of the system. The coupled equations (\ref{Kuramoto_model}) is used to develop an analytical framework for the achievement of perfect synchronization in the network. The details are described in the next section.


\section{Analytical Framework}\label{Analytical}
Here we describe the analytical framework used to derive a frequency set for achieving perfect synchronization at a targeted point in the parameter space  following the  approach reported in~\cite{skardal2014optimal, kundu2018perfect}. 
Linearization of the system ~(\ref{Kuramoto_model}) about the synchronized state ($|\theta_j-\theta_i|\rightarrow 0$) leads to the equations
\begin{widetext}
\begin{eqnarray}
\dot{\theta_i}&=&\omega_i+[-K_1 k_i^{(1)}\sin \alpha-K_2 k_i^{(2)}\sin \beta]
-K_1\cos\alpha[k_i^{(1)}\theta_i-\sum_{j=1}^NA_{ij}\theta_j]-K_2\cos\beta[k_i^{(2)}\theta_i
-\sum_{j=1}^N A_{ij}(\sum_{l=1}^N A_{jl}A_{li})\theta_j \nonumber\\
&+&\frac{1}{2}\sum_{j=1}^N A_{ji}(\sum_{l=1}^N A_{il}A_{lj}) \theta_j]~(i=1,2,\dots, N).\label{Linear_KM}
\end{eqnarray} 
\end{widetext}

The above equations can be written in vector form as, 
\begin{equation}
\label{vector_form}
 \dot{{\bf{\theta}}}=\omega+d-L{\bf{\theta}},
\end{equation}
where $d=-K_1 k^{(1)}\sin\alpha-K_2 k^{(2)}\sin\beta$ 
and $L=K_1\cos\alpha L^{(1)}+K_2\cos\beta L^{(2)}$ is the composite Laplacian consisting  of Laplacian of pairwise interaction  and  Laplacian of triadic interaction $L^{(1)}$ and $L^{(2)}$ respectively. Here, $L^{(1)}$ and $L^{(2)}$ are defined as  
\begin{eqnarray*}
L^{(1)}=D^{(1)}-A^{(1)}~\mathrm{and}~ \hspace*{.4cm} L^{(2)}=D^{(2)}-(A^{(2)}-\frac{A^{(2)T}}{2}),
\end{eqnarray*}
where $A^{(1)}$ denotes the adjacency matrix $A$, $D^{(1)}=$diag$(k_1^{(1)},k_2^{(1)},\dots,k_N^{(1)})$,
$A^{(2)}=A*(A^2)^T$, $D^{(2)}=$diag$(k_1^{(2)},k_2^{(2)},\dots,k_N^{(2)})$ with $*$ denoting the Hadamard product. Also, $k_i^{(1)}=\sum_{j=1}^NA_{ij}$ is the degree of $i$-th node considering the pairwise interaction only and $k_i^{(2)}=\frac{1}{2}\sum_{j=1}^N\sum_{l=1}^N B_{ijl}$ is the degree of $i$-th node while considering the triadic interactions.

Now, we derive the steady state solution by suitable varying the reference frame as $\omega \rightarrow \omega - \Omega$, where $\Omega$ is the group angular velocity and putting $\psi=0$. Therefore $\dot{\theta}=0$ in Eq. (\ref{vector_form}) gives the stationary points as,
\begin{equation}
\label{stationary_theta}
\theta^*=L^\dagger(\omega+d),
\end{equation}
where $L^\dagger$ is the pseudo inverse of the Laplacian $L$, which is of the form 
\begin{equation}
L^\dagger=\sum_{j=2}^N \lambda_j^{-1}v_j v_j^T,
\end{equation}
where $0=\lambda_1<\lambda_2\leq\dots \leq \lambda_N$ are the eigenvalues of $L$ and $v_j$ $(j=1,2,\dots,N)$ are the eigenvectors of $L$ with respect to the eigenvalues $\lambda_j$. 

The order parameter in Eq. (\ref{order_parameter}) then approximated as 
\begin{equation}
\label{approximated_r}
r\approx 1-\frac{\parallel \theta^* \parallel^2}{2N},
\end{equation}
where 
\begin{eqnarray*}
\parallel \theta^* \parallel^2 &=& \langle \theta^*, \theta^* \rangle \\
&=& \langle L^\dagger(\omega+d),L^\dagger(\omega+d) \rangle \\
&=& \sum_{j=2}^{N}\lambda_j^{-2} \langle v_j,\omega+d \rangle ^2.
\end{eqnarray*}
Now, by substituting the above expression in the equation (\ref{approximated_r}) we get the order parameter as
\begin{eqnarray}
\label{r_J}
r &=& 1-\frac{1}{2N}\sum_{j=2}^{N}\lambda_j^{-2} \langle v_j,\omega+d \rangle ^2 \nonumber \\
&=& 1-\frac{1}{2}J(\omega,L),
\end{eqnarray}
where $J(\omega,L)=\frac{1}{N}\sum_{j=2}^{N}\lambda_j^{-2} \langle v_j,\omega+d \rangle ^2 $ is known as synchrony alignment function (SAF). Here we note that in Eq. (\ref{r_J}), $J(\omega,L)\rightarrow 0$ limit leads to $r=1$, which denotes the perfect synchronization state. Therefore, for the achievement of perfect synchronization at a given point, we choose 
\begin{eqnarray}
\omega+d=0. \label{omega_condition}
\end{eqnarray} 
Thus the frequency set for the achievement of  perfect synchronization at a targeted point in the parameter space is given by 
\begin{equation}
\label{derived_freq}
\omega=K_1^{(p)} k^{(1)}\sin\alpha+K_2^{(p)} k^{(2)}\sin\beta, 
\end{equation}
where $K_1^{(p)}$ and $K_2^{(p)}$ are the targeted coupling strength for pairwise and triadic interaction respectively. We now proceed for the numerical verification of the proposed framework for achieving  perfect synchronization at targeted coupling strengths for fixed phase-lags $\alpha$ and $\beta$, considering different networks. 

\section{Numerical Verification} \label{Numerical}

The achievement of perfect synchronization in complex networks at a targeted point in the parameter space using the frequency set (analytically derived by the method described in the previous section) is numerically verified in this section.   For that purpose, we numerically simulate the Sakaguchi-Kuramoto model with higher order interactions on top of four different complex networks of same size ($N = 10^3$).  The considered networks are one Erd{\H{o}}s-R{\'e}nyi (ER) ~\cite{Erdos_PubMathDeb1959,Erdos_PMIHAS1960,Erdos_AMH1961} network; two scale-free(SF)~\cite{barabasi2003scale} networks with exponents $\gamma=2.5$ and $3.2$;  and one small world (SW)~\cite{watts1998collective} network.  The SF network with exponent $2.5$ is of mean degree $\langle k \rangle = 6$ and each of the other three networks have mean degree $\langle k \rangle = 8$. Numerical simulations are performed using $4$-th order Runge-Kutta (RK4) method with step size $\delta t=0.01$ for sufficiently long time after removing the transients.  For next few numerical illustrations  we take  $\alpha=\beta$ and set the target point at  $K_1^{(p)} = 0.1$ and $K_2^{(p)} = 0.5$.  In each case, we numerically continue the solutions both in forward and backward directions. 
\begin{figure}
\centering
\includegraphics[height=!,width=0.48\textwidth]{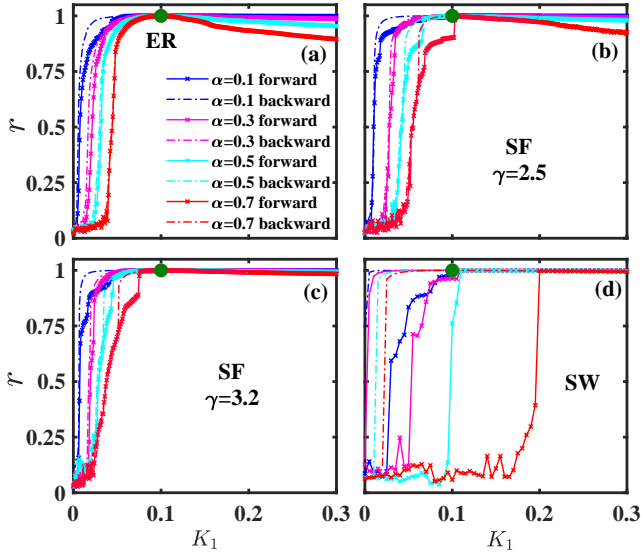}
\caption{Perfect synchronization for different networks. Numerically computed order parameter $r$ as a function of pairwise coupling strength $K_1$ shown for phase lags $\alpha$=0.1(blue), 0.3(magenta), 0.5(cyan), 0.7(red). Triadic coupling $K_2$ is set to 0.5. Green dot indicates the targeted perfect synchronization at $K_1=0.1$ with $r=1$. (a) ER network for different values of $\alpha$. $r=1$ is achieved at $K_1=0.1$ both in forward and backward simulations. (b) SF network with $\gamma=2.5$ attain $r=1$ except for forward simulations at $\alpha=0.7$. (c) SF with $\gamma=3.2$ shows perfect synchronization for all four lags in forward and backward simulations. (d) SW network achieves perfect synchronization in forward and backward simulations only for $\alpha=0.1$. For the other lags backward paths only attain $r=1$.}
\label{fig:first}
\end{figure}

\begin{figure} 
\centering
\includegraphics[height=!,width=0.48\textwidth]{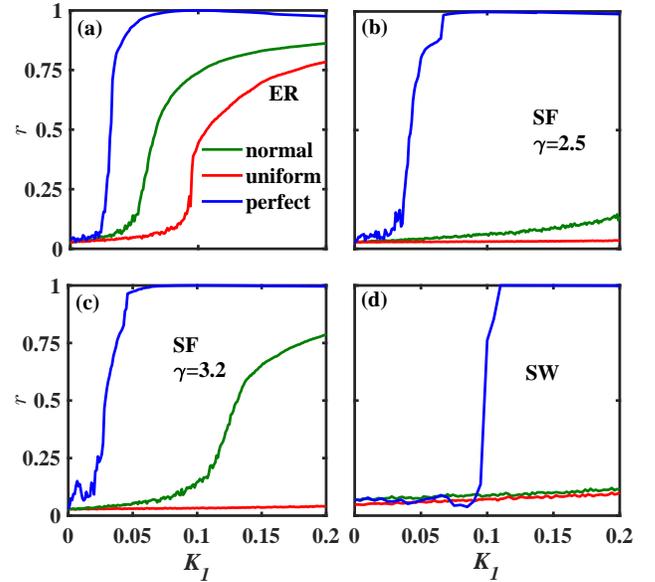}
\caption{Synchronization diagram for different frequency set. Order parameter $r$ as a function of pairwise coupling strength $K_1$. $K_2$ is fixed to 0.5 and phase lag is taken 0.5. Green, red and blue curve  is associated with the natural frequency chosen from  normal distribution, uniform distribution and derived for perfect synchronization respectively.}
\label{fig:K1_different_freq}
\end{figure}

First we fix  $K_2 = 0.5$ and vary $K_1$ around the target point $K_1^{(p)} = 0.1$ for the achievement of perfect synchronization in each of the considered networks for four different values of $\alpha = \beta = 0.1, 0.3, 0.5, \rm{and}~0.7$.  For each $\alpha$, we use the frequency set derived by the equation (\ref{derived_freq}), and obtain the statistically steady solutions  after integrating the system for long time. Although here we have chosen $\alpha = \beta$, using the derived frequency set,  perfect synchronization can be achieved at a targeted point for $\alpha\neq \beta.$ This fact will be illustrated towards the end of this section.

Starting from $K_1=0$, the solution is then continued in the forward direction by changing the value of $K_1$ in small steps and using the last values of the phase variables of the previous simulation as the initial condition for the present simulation.  In the process, we continue the solution till $K_1=0.3$ and then continue the solution similarly in the backward direction by gradually reducing the value of $K_1$ in steps of $\delta K_1=0.001$ write up to the starting value of $K_1$.  The variation of the synchronization order parameter ($r$) with $K_1$ for fixed $K_2  = 0.5$ obtained from the simulation data are shown in the FIG.~\ref{fig:first}.  From FIG.~\ref{fig:first}(a) and (c) it is observed that in case of ER and SF network with $\gamma = 3.2$,  perfect synchronization is achieved at the targeted point (marked by green dot)  during forward as well backward continuation for all values of $\alpha$. However, for other three cases (FIG.~\ref{fig:first}(b) and (d)) perfect synchronization is achieved at the targeted point during backward continuation for all considered $\alpha$, while,  for forward continuation,  it is achieved at the targeted point  for lower values of  $\alpha$.  Interesting to note here that,  for the choice of analytically derived frequency set  hysteresis is observed both for SF and SW networks and  large hysteresis loop is formed for higher $\alpha$.  This is counter intuitive,  and special result in presence of higher order interactions, because previous studies with only pairwise interactions suggest  the  negative impact of frustration parameter on the hysteresis loop~\cite{kundu2017transition}.   Therefore, from the above discussion it is apparent that using the analytically derived frequency set,  perfect synchronization can be achieved in a wide variety of networks for large values of the frustration parameter at a targeted point in the parameter space. 

\begin{figure}
\centering
\includegraphics[height=!,width=0.49\textwidth]{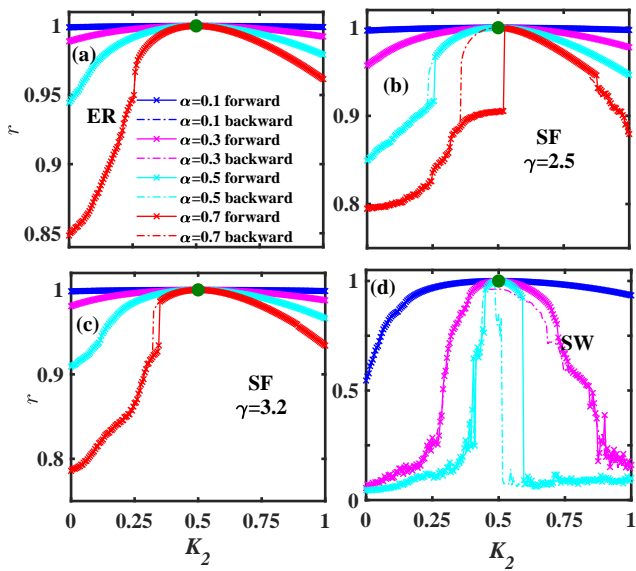}
\caption{Synchronization of SK model with HOI  for different networks. Synchronization diagrams describe the order parameter $r$ as a function of triadic coupling $K_2$ for fixed $K_1=0.1$. The blue, magenta, cyan and red curves presents the
transitions for $\alpha=$0.1,~0.3,~0.5,~0.7 respectively. The green dot denote the perfect synchronization. (a) ER network achieve $r=1$ for all four lags. (b) SF network with $\gamma=2.5$ attain $r=1$ except for $\alpha=0.7$(forward path). (c) SF network with $\gamma=3.2$ attain perfect synchronization for all $\alpha$ in both paths. (d) SW network attain $r=1$ only during forward transition(except $\alpha=0.1$).}
\label{fig:K2_v_r}
\end{figure}  

Further, the analytically derived frequency which has functional dependence both on the pairwise as well as triadic interactions not only provide perfect synchronization at a targeted point in the parameter space but also helps in achieving very high level of synchronization around the targeted point compared to other standard frequencies. FIG.~\ref{fig:K1_different_freq} shows the comparison of the level of synchronization in the considered networks for different frequency sets. Here, simulations are performed with uniform and normal frequencies along with the derived frequency set. In each case, the derived frequency set clearly provide substantially higher level of synchronization around the targeted point.

Next we fix the pairwise coupling strength $K_1 = 0.1$ and vary the triadic coupling strength $K_2$ around the targeted point ($K_2^{(p)} = 0.5$) and perform numerical simulations using the SK model on the same complex networks considered above.  As before, in this case also we numerically continue the solutions both in forward and backward directions by changing $K_2$ in the steps of $\delta{K_2} = 0.005$ for different values of $\alpha.$ The variation of $r$ with $K_2$ as obtained from the simulation data in different cases are shown in the FIG.~\ref{fig:K2_v_r} which clearly shows the impact of triadic interaction in the system.  It is observed that for all four networks perfect synchronization is achieved at the targeted coupling with different transition paths. ER and SF networks show continuous transition to synchronization for small $\alpha$. Whereas, at $\alpha=0.7$ both of the SF networks exhibit small hysteric window around the perfect synchronization point. For higher value of $\alpha$, the system (phase-lag oscillators in SW network) gradually reaches to the highest point ($r=1$), and then it sharply drops to zero. We have also observed that during the backward transition, the perfect as well as higher degree of synchronization is achieved only for small range of coupling. The reason is as follows: during the backward transition, the system is already in fully incoherent states, thus it tries to remain in the incoherent state if we decrease the coupling strength. However, when it comes closer to $ K_2^{(p)} =0.5$, the coupling strength dependent frequency drives the entire system to reach the perfect (or near perfect) synchronization regime. Further decrease of coupling strength reduces the level or degree of synchronization and drives the entire system to be fully desycnhronized. These results confirm that using the analytically derived frequency set one can achieve perfectly synchronized state either in the forward transition path or in the backward transition path.  Subsequently we analyze the analytically derived frequency set.  

\begin{figure}
\centering
\includegraphics[height=!,width=0.48\textwidth]{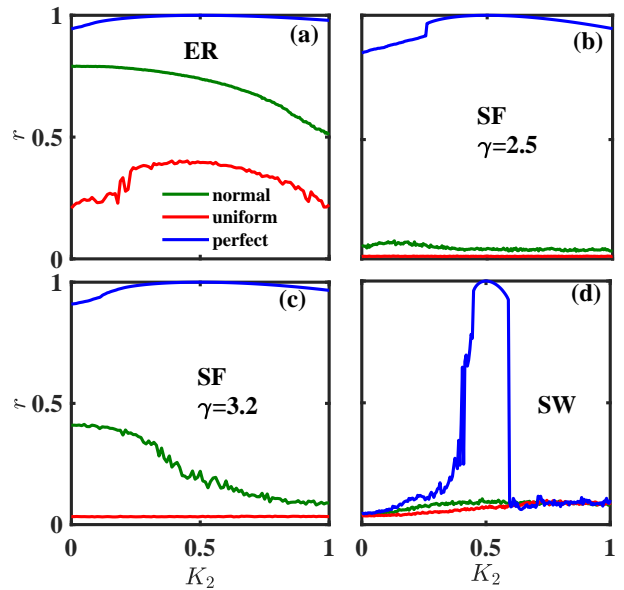}
\caption{Synchronization diagram for different frequency set for parameter $K_2$. Order parameter $r$ is plotted varying the higher order coupling strength $K_2$ for fixed $K_1=0.1$ and phase lag 0.5. The system synchronize for the derived perfect frequency set (blue) in all taken networks and  avoids synchronization in normal (green) and uniform (red) frequency distributions.}
\label{fig:K2_different_freq}
\end{figure}

At this point let us recall that the derived frequency set promotes very high level of synchronization around the targeted point as the pairwise coupling strength $K_1$ varies for fixed targeted triadic coupling strength (see FIG.~\ref{fig:K1_different_freq}). Next to check for the similar behavior of the level of synchronization we perform simulation by vary $K_2$ around the targeted point for fixed $K_1 = 0.1$. FIG.~\ref{fig:K2_different_freq} shows that in this case also the derived frequency provides very high level of synchronization compared to other frequency sets.   

\begin{figure*}
\centering
\includegraphics[height=!,width=1\textwidth]{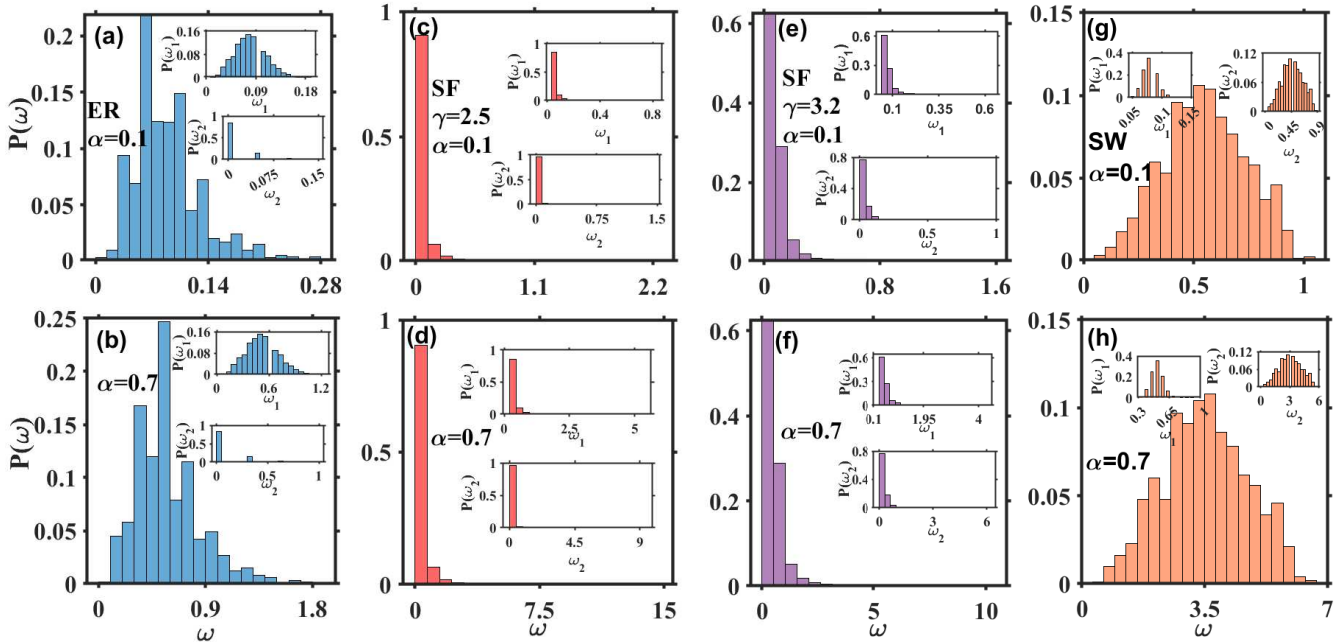}
\caption{Distribution of the frequency for perfect synchronization using the four different networks namely ER(blue), SF($\gamma=2.5$)(red), SF($\gamma=3.2$)(purple), SW(brown). In the first row the frequency is calculated for phase lag $\alpha=0.1$ and the second row the phase lag $\alpha=0.7$. We take the pairwise coupling strength $K_1^{(p)}=0.1$ and the higher order coupling strength $K_2^{(p)}=0.5$.}
\label{frequency_plot}
\end{figure*}
Now to understand the contributions of pairwise and triadic interactions in the derived frequency for achieving perfect synchronization at a targeted point, we analyze the frequency distributions corresponding to the pairwise ($\omega_1=K_1^{(p)} k^{(1)}\sin\alpha$) and triadic interactions ($\omega_2=K_2^{(p)} k^{(2)}\sin\alpha$) in the expression (\ref{derived_freq}) along with the total derived frequency distribution.   FIG. \ref{frequency_plot} shows the distribution of the derived  frequency for perfect synchronization from the equation (\ref{derived_freq})  for four different networks and two phase lag values ($0.1$ and $0.7$).  The figure also displays the distributions of $\omega_1$ and $\omega_2$ separately in the insets of each  panel.  It  is observed that the range of the frequencies  increases substantially  with lag.   As the sets $\omega_1$ and $\omega_2$ are linearly correlated with the pairwise degree $k^{(1)}$ and triadic degree $k^{(2)}$,  the distributions  also looks like the distribution of the pairwise and  triadic degrees respectively.  Since ER network have small number of triangles, from the distribution of $\omega_2$ it is clear that the effect of triangular interactions is less than the effect of pairwise interactions (FIG. \ref{frequency_plot}(a)-(b)).  For the SF networks also there are more pairwise interactions than triadic (FIG. \ref{frequency_plot}(c)-(f)).  For these networks perfect synchronization is attained at the targeted coupling up to lag $0.7$. Only in case of SW network there are  higher number of triadic interactions (FIG. \ref{frequency_plot}(g)-(h)), as a reason the effect of triadic interaction in the constructed frequency is much higher compared to ER or SF networks. This effect of triadic interaction increases the heterogeneity in the system and prevent the system to synchronize~\cite{chutani2021hysteresis}, although the predicted frequency is able to overcome this constraint and reach the global synchrony for a small vicinity of the parameter space.

\begin{figure}
\centering
\includegraphics[height=!,width=0.48\textwidth]{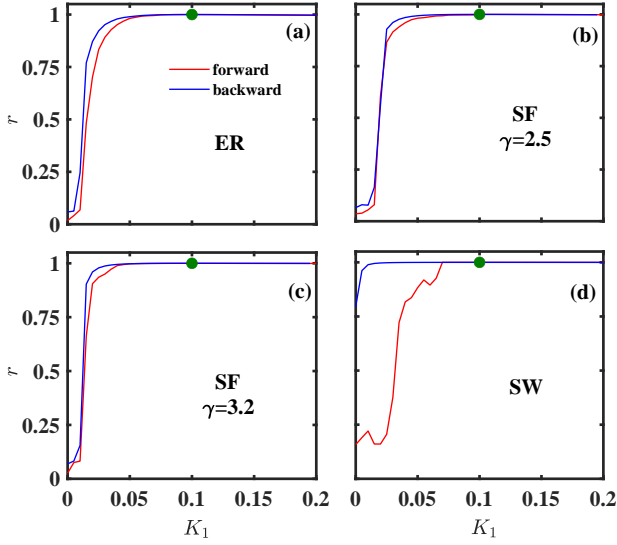}
\caption{Synchronization profile for the derived frequency set at  $\alpha=0.2$ and $\beta=0.1$. Order parameter $r$ as a function of pairwise coupling strength $K_1$. $K_2$ is fixed to 0.5. Perfect synchronization (green dot) is achieved at the targeted coupling $K_1=0.1$ for all considered networks.}
\label{fig:K1_different_alpha}
\end{figure}
\begin{figure}
\centering
\includegraphics[height=!,width=0.48\textwidth]{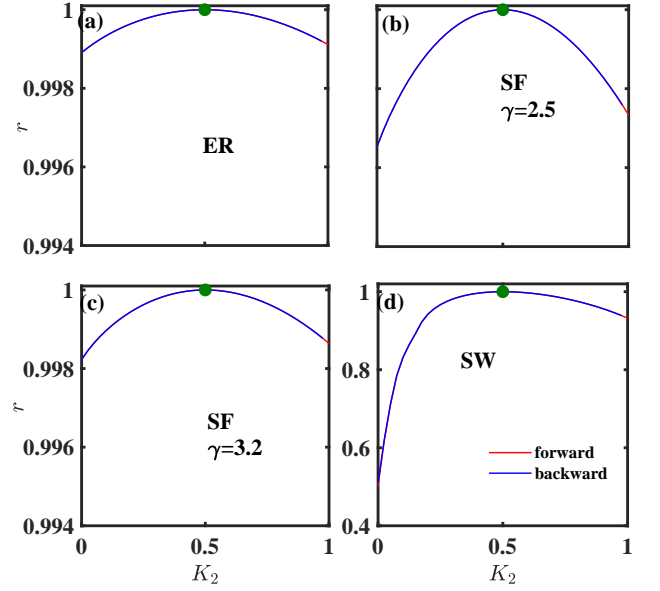}
\caption{Synchronization profile for the derived frequency set at  $\alpha=0.2$ and $\beta=0.1$. Order parameter $r$ vs $K_2$ for the fixed pairwise coupling $K_1=0.1$. All considered networks attain $r=1$ (green dot) at targeted $K_2=0.5$ both in forward(red) and backward(blue) path.}
\label{fig:K2_different_alpha}
\end{figure}

As mentioned earlier, although we have presented all the results by setting $\alpha = \beta$, the proposed scheme works in the general case where $\alpha\neq \beta.$ To check that we take $\alpha = 0.2$ and $\beta = 0.1$ and perform the numerical simulation to achieve perfect synchronization for $K_1^{(p)} = 0.1$ and $K_2^{(p)} = 0.5$ using the derived frequency set given by the Eq. (\ref{derived_freq}). FIG.~\ref{fig:K1_different_alpha} shows the variation of the order parameter computed from the simulation data around the targeted point $K_1 = 0.1$ for fixed $K_2 = 0.5$ for four considered networks. Also the variation of $r$ with $K_2$ around the targeted point for fixed $K_1 = 0.1$ for these networks are shown in FIG.~\ref{fig:K2_different_alpha}. From these two figures also the achievement of perfect synchronization at the targeted point for different phase lags associated with the pairwise and triadic interactions is quite apparent. Next we perform low dimensional reduction of the system to understand the stability of the achieved synchronization state using the derived frequency set.

\section{Low dimensional reduction}\label{lowd}
To analyze the stability of the synchronized state around the targeted parameter values we reduce the networked system in Eq.~(\ref{Kuramoto_model}) to it's low dimensional form using the collective coordinate approach~\cite{gottwald2015model, pinto2015optimal, brede2016frustration}. In this approach the phase of each oscillators is approximated by their intrinsic frequency using 
\begin{equation}
\label{approx_theta}
\theta_i(t)=\chi(t)\omega_i,
\end{equation}
where $\chi(t)$ is the collective coordinate, which is time dependent. To ensure the validity of this approach  we aim to minimize the error given by
\begin{eqnarray}
\epsilon_i(\chi) &=& \dot{\chi}\omega_i-\omega_i-K_1\sum_{j=1}^{N}A_{ij}\sin(\chi(\omega_j-\omega_i)-\alpha)-\frac{K_2}{2} \nonumber \\
&&\sum_{j=1}^{N}\sum_{l=1}^{N}B_{ijl}\sin(\chi(2\omega_j-\omega_l-\omega_i)-\beta).
\end{eqnarray}
This error will be minimum if it is orthogonal to the tangent space of the solution space given by Eq. (\ref{approx_theta}), which is spanned by $\frac{\partial \theta_i}{\partial \chi}=\omega_i$\cite{gottwald2015model, pinto2015optimal}. Now, projecting this error to the specified subspace and using the orthogonality property we obtain one-dimensional evolution equation for $\chi(t)$
\begin{equation}
\label{evo_eqn}
\frac{d\chi}{dt}=g(\chi),
\end{equation}
where $g(\chi)$ is given by
\begin{eqnarray}
g(\chi)&=&1+\frac{K_1}{\sigma^2}\sum_{i=1}^{N}\omega_i\sum_{j=1}^{N}A_{ij}\sin(\chi(\omega_j-\omega_i)-\alpha)+\frac{K_2}{2\sigma^2}   \nonumber \\
&&\sum_{i=1}^{N}\omega_i\sum_{j=1}^{N}\sum_{l=1}^{N}B_{ijl}\sin(\chi(2\omega_j-\omega_l-\omega_i)-\beta)
\label{eq.g(chi)}
\end{eqnarray}
and $\sigma^2=\sum_{i=1}^{N}\omega_i^{2}$. The one-dimensional differential Eq. (\ref{evo_eqn}) will have a stable equilibrium if $g(\chi)=0$ and $g'(\chi)<0$. At the equilibrium   $\chi$ is  independent of time and hence  Eq.(\ref{approx_theta}) indicates that all $\theta_i$'s will be time independent, which implies the system will exhibit a phase-locked solution. If such fixed point occurs at $\chi=0$, all phases will be zero, resulting perfect synchronization.  \\

\begin{figure*}
\centering
\includegraphics[height=!,width=\textwidth]{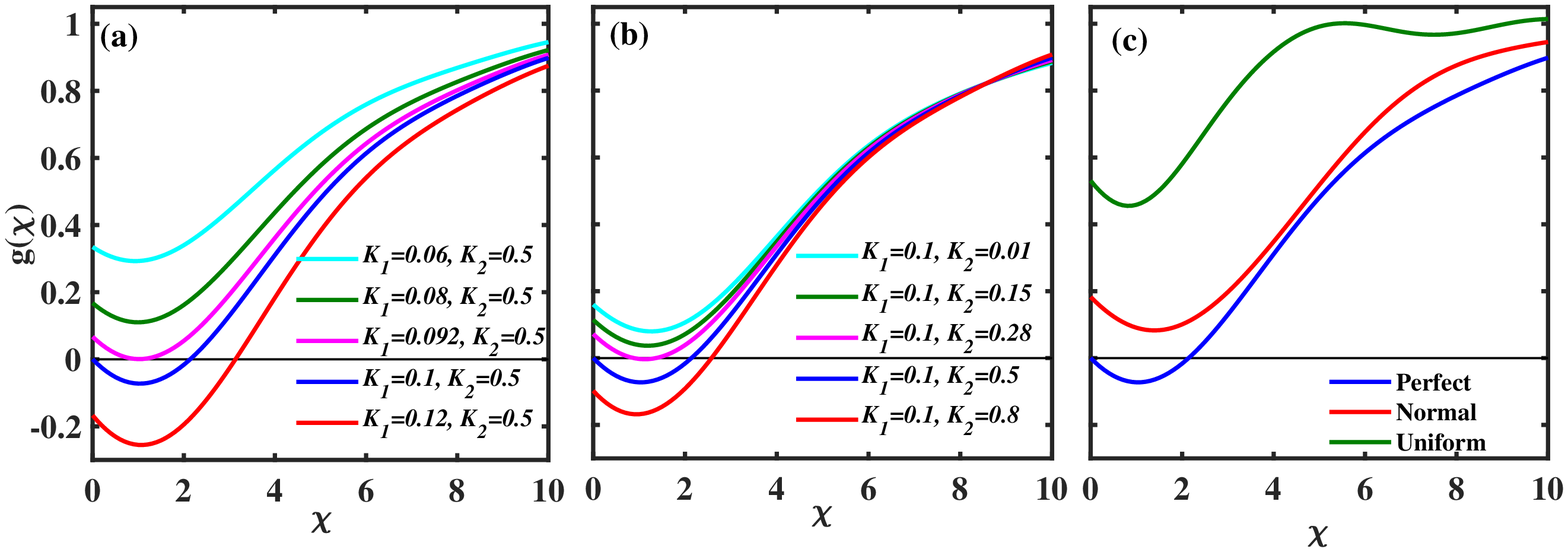}
\caption{Variation of $g$ obtained in equation(\ref{eq.g(chi)}) as a function of $\chi$ under perfect frequency for ER network. (a) Coupling $K_2$ is set to 0.5 and the value of $K_1$ are taken as 0.06(cyan), 0.08(green), 0.092(magenta), 0.1(blue), 0.12(red). (b) Coupling $K_1$ is fixed to 0.1 and the value of $K_2$ varies as 0.01(cyan), 0.15(green), 0.28(magenta), 0.5(blue), 0.8(red).(c) Derived frequency (blue), normally (red) and uniformly (green) distributed frequency is plotted, showing that only derived frequency reaches the stable synchronization state.}
\label{fig.chi_vs_g(chi)}
\end{figure*}
In FIG.~\ref{fig.chi_vs_g(chi)}(a), and \ref{fig.chi_vs_g(chi)}(b) we show the behaviour of $g(\chi)$ under derived frequency for perfect synchronization $\omega$ (targeted at $K_1=0.1,K_2=0.5$ respectively), for a set of values of $K_1$ and $K_2$ in ER network. 
We now fix $K_1$ to $0.1$ and vary $K_2$ in FIG.\ref{fig.chi_vs_g(chi)}(b). For $K_2=0.01$ and $0.15$, $g(\chi)$ does not intersect $g(\chi)=0$. Here, $g(\chi)$ touches the line $g(\chi)=0$ for the first time at $K_2=0.28$ which provides a stable phase locked solution. From this value of $K_2$ onward $g(\chi)$ crossed the line $g(\chi)=0$ with a negative slope, which indicates that system has a stable phase-locked solution (as seen in FIG.\ref{fig:K2_v_r}(a)). As per expectation the system exhibits perfect synchronization at $K_2=0.5$ ($g(\chi)=0$ and $g'(\chi)<0$ at $\chi=0$). For larger values of $K_2$ the synchronization remains stable. We do the analysis for other two networks (SF and SW) which we found very similar.\\
In 
FIG. \ref{fig.chi_vs_g(chi)}(a) we fixed $K_2$ to $0.5$ and choose $K_1$ as $0.06,~0.08,~0.092,~0.1,~0.12$. For $K_1=0.06$ and $0.08$ $g(\chi)$ does not intersect $g(\chi)=0$, i.e stable synchronization is not attended at this point. At $0.092$, $g(\chi)$ touches $g(\chi)=0$, that means stable synchronization occurs for the first time. As the $K_1$ value increases from $0.092$ the crossing point tends to zero and finally at $K_1=0.1$(targeted value), $g(\chi)$ is zero and $g'(\chi)<0$ at $\chi=0$, i.e the perfect synchronization is obtained. For the values of $K_1>0.1~(0.12)$ the solution remains stable, as we see it in FIG.\ref{fig:first}(a). \\

FIG.\ref{fig.chi_vs_g(chi)}(c) shows the relation between $\chi $ and $ g(\chi)$ for three types of natural frequency $\omega$, namely, frequency for perfect synchronization, normally distributed frequency and uniformly distributed frequency. Here the underline network is ER network. We find that the choice of $\omega$ which induce perfect synchronization, derived in Eq.(\ref{derived_freq}) gives stable equilibrium at $\chi=0$. For the other two frequency set (normal and uniform distributed) $g(\chi)$ even never intersect $g(\chi)=0$. This shows that the synchronization is stable only for the choice of frequency set derived in  Eq.(\ref{derived_freq}) among the considered set of frequencies. \\

\section{Robustness of the frequency set}\label{robustness}
The robustness of the derived frequency set ($\omega$) for perfect synchronization is now checked by adding a  perturbation to $\omega$ in the form of Gaussian noise such that $\omega_i=\omega_i+\delta \omega_i$, where $\delta\omega_i$ is drawn from a normal distribution $N(0,\sigma\omega_i)$, whose mean is $0$ and standard deviation is $\sigma\omega_i$, which represents the multiplicative noise proportional to $\omega_i$. Due to the effect of the perturbation in the optimal frequency, the system will deviate from the perfect synchronized state.  We define this deviation from synchronization state as synchronization loss and define as $\rho=1-r$, where $r$ is the synchronization order parameter defined in Eq. \ref{order_parameter}.
Now, from Eq. (\ref{approximated_r}) we get
\begin{equation}
\label{rho_expression}
\rho=\frac{1}{2N}\parallel \theta \parallel^2 \sim \frac{1}{2}{\mathrm{Var}}(\theta),
\end{equation}
where ${\mathrm{Var}}(\theta)$ denotes the variance of $\theta$. Substituting the perturbed $\omega$ in Eq. (\ref{stationary_theta}), results 
\begin{equation*}
\theta=L^\dagger\delta\omega.
\end{equation*}
 
Since $L^\dagger$ and $\delta\omega$ are approximately independent to each other, we can write 
\begin{equation}
\mathrm{Var}(\theta)=(L^\dagger)^2 \mathrm{Var}(\delta\omega)= (L^\dagger)^2 \omega^2 \sigma^2.
\end{equation}
Above equation confirms that $\mathrm{Var}(\theta)\propto \sigma^2$, where the proportion constant is $(L^\dagger)^2 \omega^2$, which is independent of $\sigma$ . Using this fact, we reach to a relation between $\mathrm{Var}(\theta)$ and $\sigma$ as $\mathrm{Var}(\theta)\sim \sigma^2$. Therefore from (\ref{rho_expression}) we get
\begin{equation}
\rho \sim \sigma^2.
\label{robust}
\end{equation}


 

\begin{figure}
\centering
\includegraphics[height=!,width=0.48\textwidth]{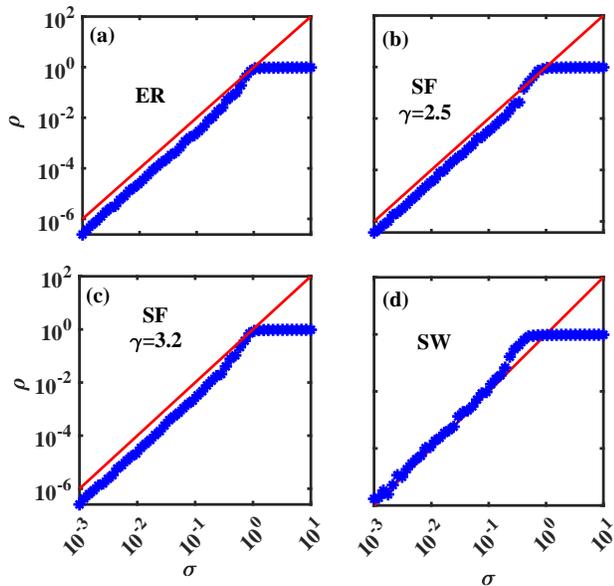}
\caption{Synchronization error$(\rho)$ under deviated frequency. We compute synchronization loss(blue) after adding noise $N(0,\sigma\omega_i)$ to the perfect frequency set. For $\sigma<1$, this error $\rho$ follows $\rho\sim\sigma^2$(blue). For $\sigma \geq 1$, synchronization fully lost, i.e $\rho \rightarrow 1$.}
\label{fig:rho_vs_sigma}
\end{figure}

A numerical verification is done in FIG.\ref{fig:rho_vs_sigma} using 4 networks, such as ER, SF with two exponents($\gamma=2.5$ and $3.2$) and SW networks. It presents $\rho$ as a function of $\sigma$. In all four cases, this figures show that for small $\sigma$, $\rho$ is small, means the deviation from synchronization is small when the deviation of the frequency from the designed frequency set is small. As the value of $\sigma$ increases (the deviation from the perfect frequency set increases), $\rho$ also increases following the rule derived in Eq. \ref{robust}. After a certain value of $\sigma$ the value of $\rho$ saturates at $1$ because at this stage  the high deviation in the frequency set lead to  a complete incoherent (desynchronized state) with $r\rightarrow 0$ which lead to $\rho\rightarrow1$. 

\section{Discussion}\label{discussion}
In this article, we developed a general mathematical framework to derive the natural frequencies of the nodes of a complex network which can ensure perfect synchronization at considerably lower coupling strength. We presented a synchrony alignment function that measures the interplay between network structure and oscillator heterogeneity and allows to get a set of frequency for perfect synchronization. Focusing on the Sakaguchi-Kuramoto model with higher order interaction, we described the effect of pairwise coupling as well as higher order coupling to reach the perfectly synchronized state at a targeted coupling strength for four different types of networks. We found that the analytically derived natural frequencies involve both structural and dynamical information of the phase frustrated Kuramoto model with higher order interaction. The synchronization is promoted by a strong alignment of the frequency vector with the most dominant Laplacian eigenvectors and the pseudo inverse operator. In all the cases, we found that, our derived frequency can help the oscillators to reach to the perfectly synchronized state where as any other frequency set does not reach perfect synchronization state even with much higher coupling strength. Later, we derive a low dimensional model to analyze the stability of the perfectly synchronized state using the collective coordinate approach. This provides a clear understanding of the stability of the synchronization state in different coupling strength and different frequency sets. We also checked the robustness of the derived frequency $\omega$ by adding a small perturbation in the form of Gaussian noise. This shows that even we add some noise in the derived frequency set, the synchronization state is not lost unless the frequency deviation is too high.
\begin{center}
\textbf{ACKNOWLEDGMENTS}\\
\end{center}
S.D. is supported by the INSPIRE program of the DST, India (Code No. IF190605).

%


\end{document}